\documentclass[12pt,]{iopart}
\expandafter\let\csname equation*\endcsname\relax
\expandafter\let\csname endequation*\endcsname\relax
\usepackage[utf8]{inputenc}
\usepackage[english]{babel}
\usepackage[T1]{fontenc}
\usepackage{dsfont}
\usepackage{amsmath}
\usepackage{amsfonts} 
\usepackage{hyperref}
\usepackage{tikz}
\tikzstyle{mynode}=[thick,draw=mygreen!30!black,fill=mygreen!25,circle,minimum size=22]
\usetikzlibrary{quantikz}

\usepackage[numbers,sort&compress]{natbib}
\begin{document}

\title[Quantum Circuit for Imputation of Missing Data]{Quantum Circuit for Imputation of Missing Data}

\author{Claudio Sanavio$^1$, Simone Tibaldi$^{2,3}$, Edoardo Tignone$^4$ and Elisa Ercolessi$^{2,3}$}

\address{$^1$ Fondazione Istituto Italiano di Tecnologia,Center for Life Nano-Neuroscience at la Sapienza, 00161 Roma, Italy}

\address{$^2$ Dipartimento di Fisica e Astronomia dell'Universit\`a di Bologna, I-40127 Bologna, Italy}
\address{$^3$ INFN, Sezione di Bologna, I-40127 Bologna, Italy}

\address{$^4$ Leithà S.r.l.~\text{\textbar} Unipol Group, Bologna, Italy}

\eads{\mailto{claudio.sanavio@iit.it}, \mailto{elisa.ercolessi@unibo.it}}

\begin{abstract}
The imputation of missing data is a common procedure in data analysis that consists in predicting missing values of incomplete data points.
In this work we analyse a variational quantum circuit for the imputation of missing data. We construct variational quantum circuits with gates complexity $\mathcal{O}(N)$ and $\mathcal{O}(N^2)$ that return the last missing bit of a binary string for a specific distribution. We train and test the performance of the algorithms on a series of datasets finding good convergence of the results. Finally, we test the circuit for generalization to unseen data. For simple systems, we are able to describe the circuit analytically, making possible to skip the tedious and unresolved problem of training the circuit with repetitive measurements. We find beforehand the optimal values of the parameters and we make use of them to construct an optimal circuit suited to the generation of truly random data. 
\end{abstract}


\maketitle

\section{Introduction}\label{sec:I}

Missing data imputation is a common task in computer science and big data analysis. In fact, datasets are often incomplete, as some of the data can have one or many entry (attribute) values that are missing. The reason for this incompleteness can either be because the data were not actually collected in the first place, or because they have been lost.
The mechanism of data loss itself is of great importance when analyzing the data. Generally, we can distinguish among three situations~\cite{Rubin1976,Rubin1978} that describe when the data are either missing at random (MAR), missing completely at random (MCAR) or missing not at random (MNAR). 

\noindent In the MAR case, the loss or presence of the data is independent of the value of the attribute, but depends on the value of other attributes. For instance, this is often the case in clinical surveys, where particular groups of people tend to omit sensible data, regardless of the values themselves.
 
\noindent In the MCAR case, the loss or presence of the data is truly random, and there is no correlation between the loss and other attributes. This process can actually present itself more often than expected, wherever communication errors or human mistakes take place. 
 
\noindent In the MNAR case, there is a correlation between the loss of an attribute and its value. For instance, this happens in a survey when the datum itself is sensible and the person does not want to reveal it.
 
\noindent In order to distinguish between the three scenarios, one should in principle know the mechanism of data loss, but most of the times this is not clear.

In the attempt of working with a complete dataset, one could simply delete the incomplete data. This procedure is discouraged, as the incomplete data can be a big portion of the collected data, and the removal of them can highly affect the analysis. 
However, even when the incomplete data are few, their instances could be of great importance for understanding the statistical properties of the dataset. For these reasons, several techniques based on statistical inference are used for the imputation of the missing data, such as maximum likelihood estimation~\cite{Murray2018,Enders2010,baker_maximum_2019}, and Bayesian inference methods applied for single and multiple imputation~\cite{Rubin1976,Rubin1978,Enders2010}.

Single imputation methods fill in the missing values of the dataset. Although this procedure is appealing, as it allows us to work with a complete dataset, it generally has the negative effect of producing biased estimates~\cite{Rubin1978} even in the MCAR case.
One example of single imputation method is the mean imputation, where missing values are filled in with the arithmetic mean of the available values of the variable. The cons of this approach is that it changes abruptly the correlations among the variables and it highly reduces the standard deviation of the dataset. 
A preferred strategy for single imputation is stochastic regression, that is able to avoid biases in the MCAR case.

\noindent However in order to deal with MCAR and MAR cases, multiple imputation methods are preferred, as they account for the deviation of the error that is brought by the introduction of an unobserved value in the dataset. 
Multiple imputation is a Bayesian inference technique that uses multiple single imputed data to generate a statistics of the missing values. We have to point out that multiple imputation has not the goal to impute the missing value~\cite{Glas2010}, but instead it aims to correctly use the incomplete data for extrapolating information on the complete dataset statistics. 

Nevertheless in many cases of relevance, single imputation is what is needed. This is the case for image inpainting, a subclass of data imputation problems where the goal is to fill holes in images or videos. 
In this field a parallel approach has been provided by the field of machine learning (ML) which introduced new techniques such as generative neural networks. The ML approach has been applied successfully both for image inpainting problems~\cite{Yoon2018,Elharrouss2020,miao_itrans_2024} and for single imputation of more complicated datasets~\cite{Wang2022}. 

In the wake of the enormous success of ML algorithms, the new field of quantum machine learning has born in recent years~\cite{Benedetti2019a}, with the hope of bringing together the versatility of ML algorithms with the great expectations lying in quantum computing~\cite{NielsenChuang2010}. 

In this work we tackle the problem of estimating the missing value of a datum using quantum variational algorithms. These algorithms belong to the field
of quantum machine learning (QML).

A famous QML algorithm is the Quantum Circuit Born Machine (QCBM), that uses the Born rule to generate a target distribution.
In particular, the QCBM~\cite{Benedetti2019} with $N$ qubits is a quantum variational circuit that takes as input the state $|0\rangle^{\otimes N}$, evolves it with the parameter dependent unitary operator $\hat{U}(\Theta)$ and returns the state $|\psi(\Theta)\rangle=\hat{U}(\Theta)|0\rangle^{\otimes N}$. The $M$ parameters $\Theta=(\theta_1,\dots,\theta_M)$ are selected in order to minimize a chosen cost function, such as the distance between the frequency distribution of the measured output states and the target probability distribution.
The optimization of parameters involves updating their values based on multiple measurements of the output state. This process poses a common challenge in optimization procedures. On the one hand, a significant number of measurements must be taken to encompass all potential measurement outcomes, typically on the order of $2^N$, with some exception for particular distributions~\cite{Consiglio2023}. On the other hand, determining the optimal update for the parameters is not straightforward, as quantum circuits often exhibit extensive regions in parameter space where the cost function remains essentially constant, the well-known problem of Barren Plateaus (BPs)~\cite{Mcclean2018}. 
Consequently, it is a hard problem to find the best choice of the parameters. 

Inspired by the general setup of the QCBM, we define a quantum circuit dedicated to the imputation of missing data, the Quantum Imputation Circuit (QIC). Our analysis of the QIC tries to solve the two aforementioned problems.  

In Sec.~\ref{sec:II} we describe our circuit, the QIC, that we use to impute the missing data. In Sec.~\ref{sec:III} we show the results obtained on several datasets with different distributions. In Sec.~\ref{sec:IV} we test the ability of the QIC to generalize the imputation to instances that it has not seen during the training. 
Finally Sec.~\ref{sec:V} draws the conclusion and the outlooks of the work. 

\begin{figure}[t]
  \centering
\begin{quantikz}
\lstick{$\ket{0}$} &\gate{H}\gategroup[2,steps=1,style={dashed,rounded corners,inner xsep=2pt},background]{}              
&\gategroup[3,steps=5,style={inner sep=1pt},background]{{\sc linear}}    \qw                &\qw      &\qw                &\ctrl{2}&\qw&  \ctrl{1}&\qw               &\gategroup[3,steps=-2,style={inner sep=1pt},background]{{\sc quadratic}}\qw\\
\lstick{$\ket{0}$}&\gate{H}
&\qw                &\ctrl{1} &\qw                &\qw	   &\qw                 &\ctrl{1}&\qw	           & \qw{}\\
\lstick{$\ket{0}$}&\qw
&\gate{R_y(\alpha_0)}&\targ{} &\gate{R_y(\alpha_1)}&\targ{}&\gate{R_y(\alpha_2)}&\targ{} &\gate{R_y(\alpha_3)}& \meter{}
\end{quantikz}  \caption{The Quantum Imputation Circuit (QIC) for the case of $N=2$ input qubits and 1 output qubit. The \text{dashed} box is used during the optimization, to construct the superposition of all the possible inputs. The \text{LINEAR} box shows the linear circuit, and the \text{QUADRATIC} box shows the additional part that constitutes the quadratic circuit. Finally, only the output qubit needs to be measured.}
  \label{fig:figure1}
\end{figure}
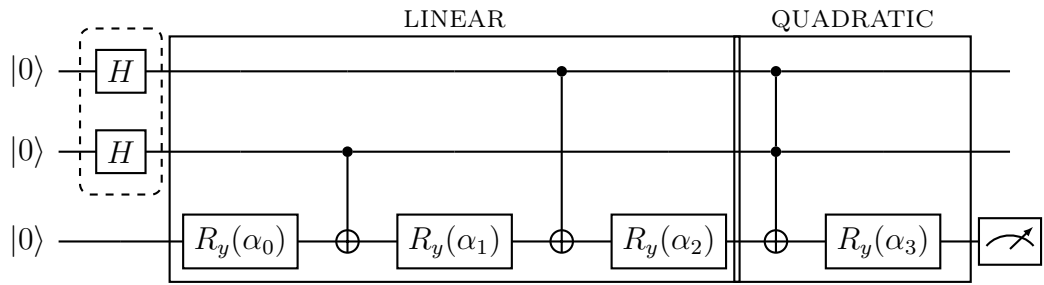

\section{Results}
\subsection{The Quantum Imputation Circuit}\label{sec:II}

We describe now the proposed QIC for the imputation of missing data, that is shown in Fig.~\ref{fig:figure1}.

Suppose our data is a collection of bit-strings $X\in\{0,1\}^{N+1}$ that are composed by $N+1$ binary variables and follow a certain probability distribution $p(X)$. This is our complete dataset that we will use for the training. 
Suppose now that we want that our machine returns the value of the $(N+1)$-th bit when the values of all the other $N$ bits are given, reproducing the probability distribution $p(X)$.

The QIC acts on the target qubit, initialized in $|0\rangle,$ and modifies its state depending on the values of the other qubits.
For any input state $|n\rangle|0\rangle$, where $|n\rangle$ is a binary representation with $N$ bits of the number $n$, the output of the circuit is 
\begin{equation}\label{eq:state}
|n\rangle(\cos\theta_{n}|0\rangle+\sin\theta_{n}|1\rangle).
\end{equation}

\noindent The probability that the target qubit is set to $0$ or $1$ is conditioned by the dataset as
\begin{eqnarray}\label{eq:prob01}
p(0|n)=&\cos^2\theta_{n},\nonumber\\
p(1|n)=&\sin^2\theta_{n}.
\end{eqnarray}
 
\noindent In order to represent any possible probability distribution we would need a parameters vector $\Theta=(\theta_0,\dots,\theta_{2^N-1})$ of dimension $2^N$. Since the circuits applies a transformation only on the output qubit and uses the input qubits as control, the unitary transformation $U(\Theta)$ that represents the QIC expressed in the computational basis is a block diagonal matrix 

\begin{equation}\label{eq:unitaryCQBM}
U(\Theta)=\bigoplus_{n=0}^{2^N-1}R_y(\theta_n),
\end{equation}
 
\noindent where $R_y$ is the rotation matrix

\begin{equation}\label{eq:rotation}
R_y(\theta_n) = 
\begin{pmatrix}
\cos\theta_n & -\sin\theta_n \\
\sin\theta_n & \cos\theta_n
\end{pmatrix}.
\end{equation}

\noindent From the explicit form of the unitary operator is clear that we need to give an independent value to all the $2^{N}$ parameters. 
$U(\Theta)$ can be reproduced using a sequence of multiple qubits controlled gates C$_{m}$NOT, with $m=1,\dots,N$, where $m$ qubits are used as control and the $(N+1)$-th qubit is the target. Each $l$-th control gate is preceded by a rotation $R_y(\alpha_l)$ applied on the target qubit. The combination of the $y$ rotations and the CNOTs leads to a global unitary that has the form expressed in Eq.~\eqref{eq:unitaryCQBM} as it will be shown in the next section. 

\noindent The number of ways we can place the C$_{m}$NOT gates using the $N$ input qubits is 
$\begin{pmatrix}
N\\
m
\end{pmatrix}$. Using all the possible combination of C$_{m}$NOT gates we have a total number of gates that is

\begin{equation}
\sum_{m=1}^N\begin{pmatrix}
N\\
m
\end{pmatrix} =2^N-1,
\end{equation}

\noindent that corresponds to $2^N-1$ parameters, as each of the control gates is followed by a parametric rotation $R_y$. In order to recover the $2^N$ needed parameters, we apply an extra parametric rotation on the target qubit. Therefore, our circuit starts and ends with a parametric rotation.

Clearly, this circuit is able to reproduce any possible data distribution but it has the negative feature of requiring an exponential number of gates. Hence, to improve the feasibility of the circuit we restrict our set of gates to only CNOTs for reproducing the dataset distribution. The number of controlled gates (and consequentially of parameters) reduces to

\begin{equation}\label{eq:linearM}
M^{\text{lin}} = \sum_{m=0}^1\begin{pmatrix}
N\\
m
\end{pmatrix} =N+1,
\end{equation}

\noindent that is linear with $N$. We call this ansatz the \textit{linear} QIC.

If we introduce also the C$_2$NOTs, the number of controlled gates scales quadratically with $N$, as

\begin{equation}\label{eq:quadraticM}
M^{\text{qua}} = \sum_{m=0}^2\begin{pmatrix}
N\\
m
\end{pmatrix} =\frac{N^2+N+2}{2}.
\end{equation}

\noindent We call this ansatz the \textit{quadratic} QIC.

Note that the term with $m=0$ in the summations of Eqs.~\eqref{eq:linearM} and \eqref{eq:quadraticM} accounts for the extra final rotation. 

In theory, following the same procedure, we could extend the circuits by adding multi-controlled gates acting on a larger number of qubits (C$_3$NOT, C$_4$NOT,$\dots$) followed by parametric rotations on the target qubit, up to a maximum number $2^N$. 

We call this the {\it{exponential}} circuit which, having an exponential number of parameters, can attain a perfect reproduction of the original dataset~\cite{Barenco1995}. Clearly, the growing number of CNOTs makes both the circuit construction and the parameters optimization unfeasible. For these reason, in this work we only analyzed the linear and quadratic circuits which shows to have good performance with less parameters.

\subsection{Analytical description of the output state}\label{sec:analytical_description}

Since the QIC has a well defined structure, we can recover the analytical expression of each $\theta_n$ angle expressed in Eq.\eqref{eq:unitaryCQBM} as a function of the rotations $R_y(\alpha_i)$, with $i=1,\dots,M$ present in the circuit. In fact, we can pull all the $R_y$ rotations at the beginning of the circuit and collect them into a single rotation operator.

Starting with the linear circuit, there are in total $N+1$ rotation angles $\alpha_i$ and $N + 1$ qubits $q_n$ with $n = 0, \dots, N$, $q_0$ being the imputation qubit.
Thanks to this particular topology the circuit's unitary matrix is in a block-diagonal form. There are $N$ blocks where each is a $(2\times 2)$ matrix $U_{\boldsymbol{b}}$, that form the total matrix:
\begin{equation}
\label{eq:unitary}
    U = \bigoplus_{\boldsymbol{b}} U_{\boldsymbol{b}}, \quad \boldsymbol{b} = \{b_1, \dots, b_N\} \in \{0,1\}^{N}
\end{equation}
where $\boldsymbol{b}$ is every possible input bitstring. To understand the shape of each $U_{\boldsymbol{b}}$ we can simplify the circuit structure by commuting all the $R_y$ rotations to the beginning of the circuit. Let us start with an example: move the second rotation $R_y(\alpha_1)$ to the left of the first CNOT (check Fig.~\ref{fig:figure1}). This means:

\begin{eqnarray}
\label{eq:commute}
R_y(\alpha_1)_{q_0} \text{CNOT}_{q_1, q_0} &=& R_y(\alpha_1)_{q_0} [\ket{0}\bra{0}_{q_1} \otimes \mathds{1}_{q_0} + \ket{1}\bra{1}_{q_1} \otimes X_{q_0}] \nonumber \\ 
    &=&[\ket{0}\bra{0}_{q_1} \otimes R_y(\alpha_1)_{q_0} + \ket{1}\bra{1}_{q_1} \otimes X_{q_0}R_y(-\alpha_1)_{q_0}] \nonumber\\
    &=&\text{CNOT}_{q_1, q_0} R_y[(-1)^{b_1}\alpha_1]_{q_0}
\end{eqnarray}

\noindent where $b_1=0 $ (or 1) if qubit $q_1$ is in state $\ket{0}$ (or $\ket{1}$). This commutation tells us that if the control qubit $q_1$ is in $\ket{1}$ the gate $R_y(\alpha_1)$ becomes $R_y(-\alpha_1)$ (because it anti-commutes with $X$). Now both $R_y(\alpha_0)$ and $R_y(\alpha_1)$ are at the beginning of the circuit and we can combine them into one rotation:
\begin{equation}
R_y[(-1)^{b_1}\alpha_1]R_y(\alpha_0) = R_y[\alpha_0 + (-1)^{b_1} \alpha_1]
\end{equation}
Now, by commuting all the rotations at the beginning of the circuit we have a simpler structure: a single initial rotation $R_y(\theta_{\boldsymbol{q}})$ followed by all the CNOTs.
The total angle of rotation $\theta_{\boldsymbol{b}}$ is given by:
\begin{eqnarray}
    \theta_{\boldsymbol{b}} &= \alpha_0 + (-1)^{b_1} \alpha_1 + (-1)^{b_1 + b_2} \alpha_2 + \dots + (-1)^{b_1 + b_2 + \dots b_{N}} \alpha_N \nonumber\\
    &= \alpha_0 + \sum_{n=1}^N (-1)^{\sum_{j = 1}^{n} b_j} \alpha_n
\end{eqnarray}
while the action of all the CNOTs introduces a possible $X$ rotation conditioned on the bitstring values:
$$
X^{b_1 + \dots + b_n}
$$
where all the sums are intended modulo 2. In this way it is easy to obtain a relation for the matrices $U_{\boldsymbol{b}}$ of eq.~\eqref{eq:unitary} by defining the bitstring-dependent partial sum up to qubit $n$:
$$
S_n (\boldsymbol{b})= \sum_{i=1}^n b_i \mod 2
$$
where the sum runs from $i=1$ because only the input qubits can affect this sum and we set by definition $S_0(\boldsymbol{b}) \equiv 0$.
For instance, if the bitstring of input qubits is $\boldsymbol{b} = 101$ then:
\begin{eqnarray}
    S_0(101)&\equiv&0 \nonumber\\
    S_1(101)&=&1 \nonumber\\
    S_2(101)&=&1 \oplus 0 = 1 \nonumber\\
    S_3(101)&=&S_N = 1 \oplus 0 \oplus 1 = 0.\nonumber
\end{eqnarray}
Consequently we can write $\theta_{\boldsymbol{b}} = \sum_{n=0}^N (-1)^{S_n} \alpha_n.$ Then, for every possible bitstring of the input qubits we have the corresponding $2 \times 2$ matrix:
\begin{equation}
\label{eq:blocks}
U_{\boldsymbol{b}} = X^{S_N} R_y(\theta_{\boldsymbol{b}}).
\end{equation}

\noindent Note that the explicit expression of $U_{\boldsymbol{b}}$ can be related to the general formula of Eq.~\eqref{eq:unitaryCQBM}, since if $S_N = 0$, then $U_{\boldsymbol{b}} = R_y(\theta_{\boldsymbol{b}})$ and if $S_N = 1$, then $U_{\boldsymbol{b}} = R_y(\theta_{\boldsymbol{b}}+\pi/2)$.

For the quadratic circuit we can repeat the similar steps and obtain the analytical form of the circuit function:

\begin{equation}
    U_{\boldsymbol{q}} = X^{S_N + Q_{N-1, N}} R_y \left(\sum_{n=1}^N \boldsymbol{\theta}_n\right)
\end{equation}
where $Q_N, \boldsymbol{\theta}_n$ and other details of this derivation are in~\ref{sec:appII} along with the exponential circuit analytical form.

\subsection{Limits of the QIC}

Our purpose is to use the QIC to reproduce the target state 
\begin{equation}
\ket{\phi_T}=\frac{1}{\sqrt{2^N}}\sum_{a=\{0,1\}}\sum_{n=0}^{2^N-1}\sqrt{p(a|n)}|n\rangle|a\rangle,
\end{equation}

\noindent where the conditioned probability $p(a|n)$ are defined in Eq.\eqref{eq:prob01}.

\noindent When a number $M\leq2^N$ of parametrized rotations is used in the circuit, the output state $|\psi(\bar{\Theta})\rangle$ is a function of the $2^N$ dimensional parameters vector $\bar{\Theta}=(\bar{\theta}_0,\bar{\theta}_1,\dots)$,

\begin{equation}
|\psi(\bar{\Theta})\rangle=\frac{1}{\sqrt{2^N}}\sum_{n=0}^{2^N-1}(\cos\bar{\theta}_n|n\rangle|0\rangle+\sin\bar{\theta}_n|n\rangle|1\rangle),
\end{equation}

\noindent where only $M$ of the angles $\bar{\theta}_n$ are linearly independent.

A common measure that accounts for the similarity between two distributions $p$ and $q$ is the Hellinger distance~\cite{jeffreys_invariant_1997}, defined as

\begin{equation}\label{eq:Hellinger}
d_H(p,q) = \sqrt{1-\sum_x\sqrt{p_xq_x}},
\end{equation}

\noindent where the term $\sum_x\sqrt{p_xq_x}$ is known as Battacharrya coefficient. The Hellinger distance is chosen for its simplicity, and for its property of being a monotone function of other commonly used distances between distributions, such as the Jensen--Shannon distance and the Battacharrya distance~\cite{saez_comparative_2013}. 
We can calculate this value between the target distribution represented by the state $|\phi_T\rangle$ and the output state as

\begin{eqnarray}\label{eq:tracedistance}
d_H({\phi_T},\psi(\bar{\Theta}))&=\sqrt{1-|\langle{\phi_T}|{\psi(\bar{\Theta})}\rangle|}\\
&=\sqrt{1-\bigg{(}\sum_{n=0}^{2^{N}-1}\frac{\cos(\theta_n-\bar{\theta}_n)}{2^{N}}\bigg{)}},\nonumber
\end{eqnarray}

\noindent with $-\frac{\pi}{2}\leq\theta_n-\bar{\theta}_n\leq\frac{\pi}{2}$. In order to understand what is the error that we could get in reproducing $|\phi_T\rangle$ using only $M$ parameters, we calculate the maximum distance we can get between all the possible distributions $|\phi_T\rangle$ and the output state when the rotation parameters $\bar\Theta$ are optimized, namely

\begin{equation}
\max_{|\phi_T\rangle}\min_{\bar{\Theta}}d_H(\phi_T,\psi(\bar{\Theta})).\nonumber
\end{equation}

The minimization of the distance is provided by the ansatz $|\tilde{\phi}\rangle$, that has without loss of generality the first $M$ parameters equal to the correspondent target angles $\bar{\theta}_i=\tilde{\theta}_i=\theta_i$. With this choice we can write the minimum of the distance as 

\begin{eqnarray}
\min_{\bar\Theta}&d_H(\phi_T,\psi(\bar{\Theta}))=d_H(\phi_T,\tilde{\phi})\\
&=\sqrt{1-\bigg{(}\frac{M}{2^{N}}+\sum_{i=M+1}^{2^{N}}\frac{\cos(\theta_i-\bar{\theta}_i)}{2^{N}}\bigg{)}}.\nonumber
\end{eqnarray}

\noindent The resulting $2^N-M$ angles are linearly dependent on the first $M$ fixed angles. Hence, the ansatz does not ensure that the other parameters can be made equal to their correspondent targets. The maximum value of the Hellinger distance among all the possible distribution represents the maximum error we can get when we have optimized over the $M$ parameters. It is obtained when the remaining angles are $\theta_i=\bar{\theta}_i+k\frac{\pi}{2}$, with $i=M+1,\dots,2^N$, and $k\in\mathbb{Z}$. This yields

\begin{equation}\label{eq:upperbound}
\max_{|\phi_T\rangle}\min_{\bar{\Theta}}d_H(\phi_T,\bar{\phi})=\sqrt{1-\frac{M}{2^{N}}}.
\end{equation}

\noindent The ansatz $|\tilde{\phi}\rangle$ represents a circuit that has learned perfectly a subset of $M$ parameters. Eq.~\eqref{eq:upperbound} provides an upper bound of the Hellinger distance between $|\phi_T\rangle$ and $|\psi(\Theta)\rangle$ if the optimization over the rotation parameters vector $\Theta$ was successful. 

Thus, the upper bound we have found in Eq.~\eqref{eq:upperbound} tells us what is the maximum error we can get when we use the QIC. 

\subsection{Parameter optimization}

One of the benefits of using the QIC for imputation and reproduction of conditional distribution probabilities is given by the optimization of the parameters. Usually, a quantum variational circuit, in particular the QCBM, has to deal with a series of measurements and subsequent updates of parameters. However, this procedure presents many problems. First, in order to get some significant statistics of the generated distribution we need to take an exponential amount of measurements, as an $N$ qubits circuit has $2^N$ potential outcomes. Second, the optimization process of the circuit is itself problematic, as it has to deal with BPs~\cite{Mcclean2018} which represent a significant portion of the parameter space where the cost function remains flat, thereby offering little guidance on how to proceed with the optimization process. This makes gradient-based methods utterly non-efficient, but it poses a fundamental challenge also to the application of gradient-free methods, such as the Bayesian optimizations strategy. 
In~\ref{sec:appI}, we provide evidence for the presence of BPs within our circuit by analyzing its entanglement entropy~\cite{Cerezo2021}. Additionally, we demonstrate the limitations of standard parameter measurement and update procedures in the context of BPs.  While we attempted to employ Bayesian optimization, its effectiveness was ultimately compromised by the plateaus, preventing us from successfully minimize the cost function. We have included these findings in~\ref{sec:appI} to distinguish our proposed method from previous approaches.

In the QIC in fact, the optimal parameters can be found just solving a constrained problem. In order to reproduce the results, the target distribution has to be such that for each input value $n$, $p(n,0)+p(n,1)=1/\sqrt{2^N}$. This is a required pre-processing that we need to do on the dataset. Finally the optimal parameters are chosen such that the Hellinger distance \eqref{eq:Hellinger} is minimal, where $\bar{\Theta}_n=\arccos\sqrt{p(0|n)}$. The analytical description of the circuit we have described in Sec.~\ref{sec:analytical_description} and detailed in~\ref{sec:appII} allows us to efficiently solve the optimization problem. 

\subsection{Imputation of probability distributions}\label{sec:III}

In this section we test the variational circuit for imputation of missing data. We create several ad-hoc datasets where the probability of the state $|n\rangle|0\rangle$ follows different distributions listed here.

\begin{figure}
\centering
 \includegraphics[width=0.8\textwidth]{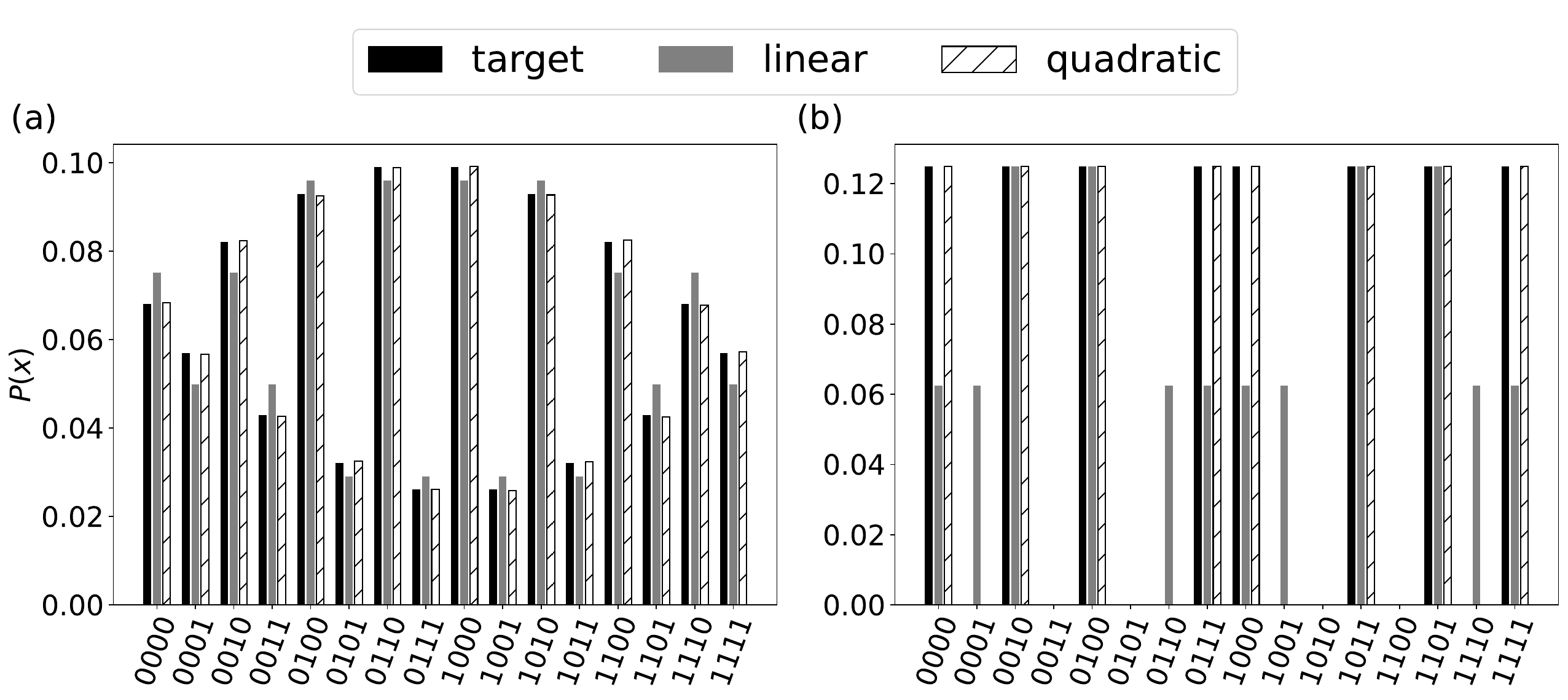}
\caption{ (a) The Gaussian distribution of Eq.~\eqref{eq:Gaussian}  and the respective output of the linear and quadratic QICs.  (b) The majority distribution of Eq.~\eqref{eq:majority}  and the respective output of the linear and the quadratic QICs. The plots show a number of input qubits $N=3$ and represent the output of the circuit after the optimization of the parameters. The linear circuit is able to reproduce well the main features of the Gaussian distribution, whereas the quadratic circuit is needed to reproduce the majority distribution.} 
   \label{fig:figure3}
\end{figure}

\paragraph{Gaussian distribution.} The Gaussian like distribution is defined as 
\begin{eqnarray}\label{eq:Gaussian}
p(n,0)&=\frac{1}{\sqrt{2^{N}}}\frac{1}{\sqrt{2\pi}}e^{-(n-(N-1)/2)^2},\nonumber\\
p(n,1)&=\frac{1}{\sqrt{2^{N}}}(1-p(n,0)).
\end{eqnarray}
The distribution can be seen in Fig.~\ref{fig:figure3}(a) (dark histogram).

\paragraph{Majority distribution.}  The majority distribution shown in Fig.~\ref{fig:figure3}(b) (dark histogram) assigns to the target qubit the value which corresponds to the most frequent value in the input. We define the function $f_x(n)$ that gives the frequency of the bit $x$ in the binary representation of the number $n$. With $f_x(n)$ we create a probability distribution in which each bitstring has probability:

\begin{equation}\label{eq:majority}
p(n,x)=
\begin{cases}
\frac{1}{\sqrt{2^{N}}}&\text{if } f_x(n)>f_{\bar{x}}(n) ,\\
\frac{1}{2\sqrt{2^{N}}}&\text{if } f_x(n)=f_{\bar{x}}(n),\\
0\qquad&\text{otherwise},
\end{cases}
\end{equation}

\noindent with $\bar{x} = x\oplus 1$. 

\begin{figure}[ht!]
\centering
 \includegraphics[width=.31\textwidth]{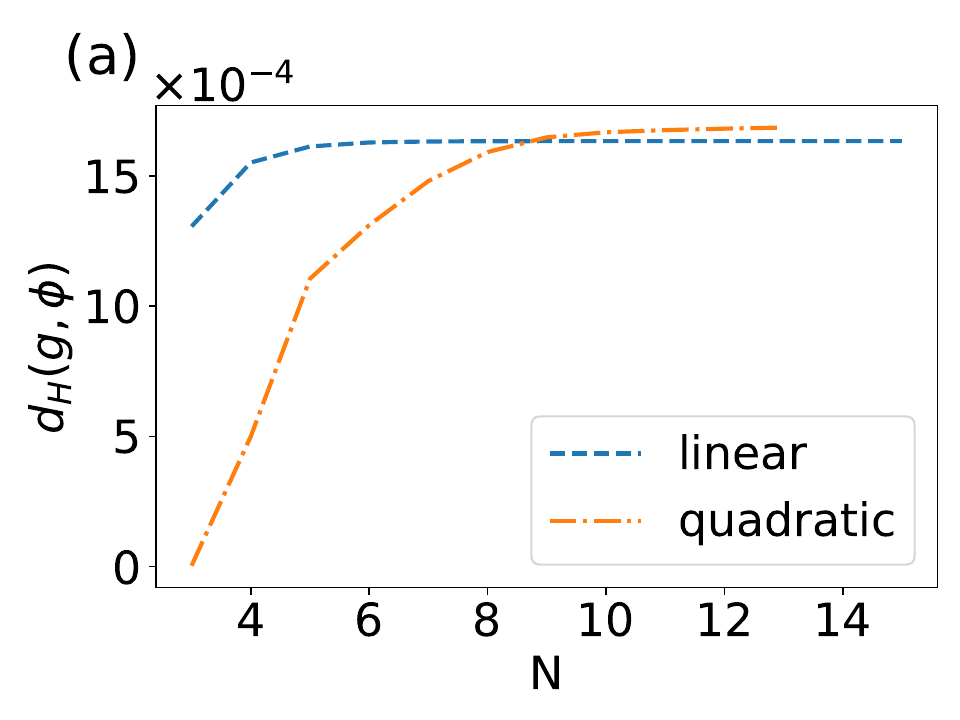}
 \includegraphics[width=.31\textwidth]{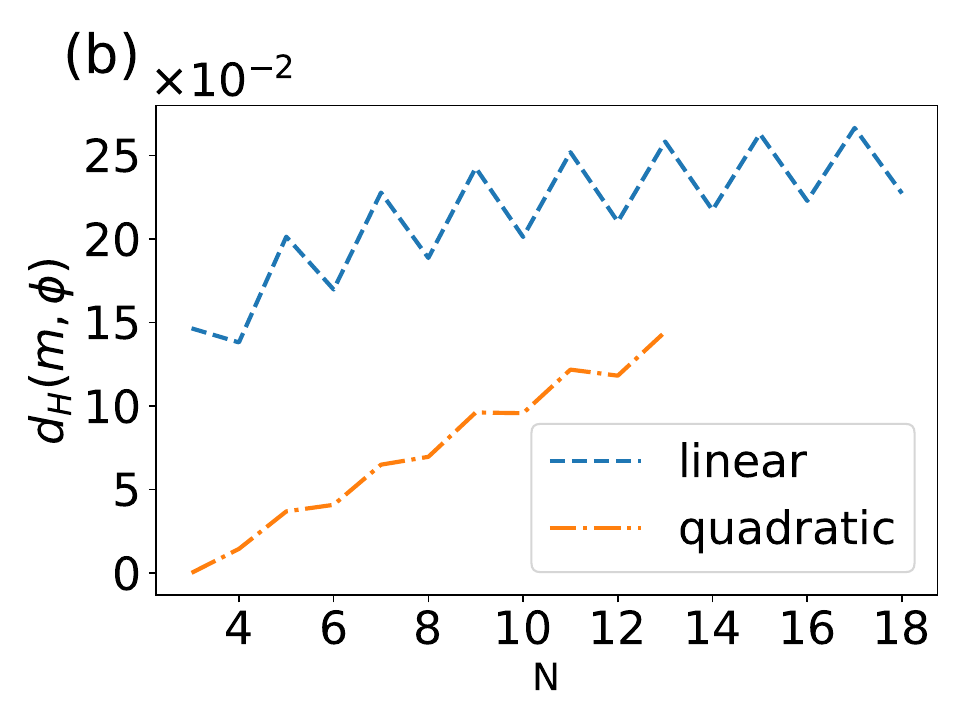}
  \includegraphics[width=.31\textwidth]{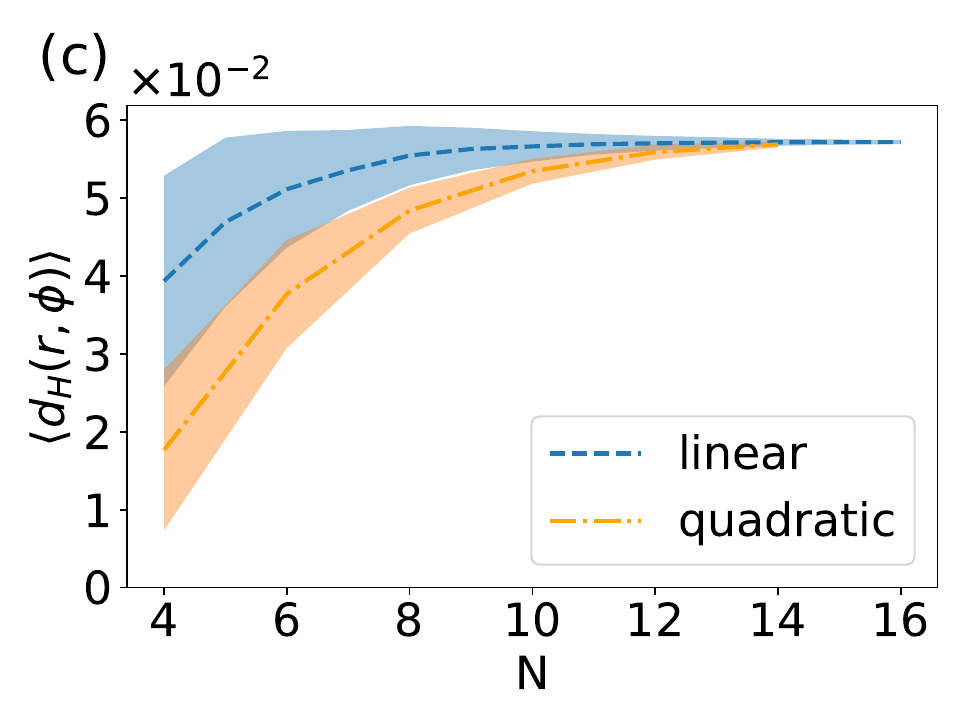}
\caption{The Hellinger distance between the target distribution and the output distribution of the optimized circuit. In (a) the target is the Gaussian distributions $g$ of Eq.~\eqref{eq:Gaussian}, and in  (b) the target is the majority distribution $m$ of Eq.~\eqref{eq:majority} .  In (c) the mean value of $d_H$ obtained for 100 random distributions for each number of input qubits $N$. The coloured area represents the variance obtained from the different random distributions.} 
   \label{fig:figure4}
\end{figure}

\subsubsection{Numerical simulations}
We have tested the performance of both the linear circuit and the quadratic circuit on either the Gaussian or the majority distribution, using the \text{SLSQP} classical optimizer to minimize expression~\eqref{eq:tracedistance}. From Fig.~\ref{fig:figure3} we can grasp how the target distributions are reproduced by the circuit. For example, in Fig.~\ref{fig:figure3}(a) both the linear and quadratic QIC are successful at reconstructing the Gaussian distribution while in Fig.~\ref{fig:figure3}(b) we see that the linear circuit was not able to reproduce the majority distribution, while the quadratic circuit could reach a much more similar output. The similarity between the distributions is given by the Hellinger distance defined in Eq.~\eqref{eq:Hellinger} and it is plotted in Fig.~\ref{fig:figure4} for the Gaussian and the majority distributions as a function of the number of input qubits. 

\noindent In the case of the Gaussian distribution, the linear and the quadratic circuit give a similar output. The quadratic circuit gives a worse result than the linear circuit for $N>10$ because of optimization errors.  Both the distances reach a plateau for large enough $N$. 
In the case of the majority distribution we see a great improvement when using the quadratic QIC.  
We see that the circuit is more prominent to reproduce the distribution for even values of $N$, that in the case of a same number of 0 and 1 gives an equal probability to assign 0 or 1 to the output qubit. 

In Fig. ~\ref{fig:figure4} (c) we plot the mean value of $d_H$ and its standard deviation obtained optimizing the circuit for 100 random distributions for each number of input qubits. Generally, both the linear and quadratic QICs reach a plateau value of the Hellinger distance of less than $0.06$ for large $N$.

\subsection{Does the QIC generalize?}\label{sec:IV}

By generalization we mean the ability of the parametric circuit to correctly complete the input data that were not present in the training set.
The question whether a QCBM does generalize has already been posed in the literature~\cite{Gili2023,Benedetti2019,Coyle2021}, and the answer reflects the fact that the QCBM takes as input a vector where all qubits are set to $|0\rangle$, and applies a global unitary transformation to reproduce the dataset distribution. As a consequence, the generalization to unseen data is possible only if the cost function during training does not go to zero, an event that would signify mere memorization of the available data. This behaviour is related to the expressibility~\cite{Sim2019,Hubregtsen2021} of the circuit, that accounts for how much of the Hilbert space is spanned by the circuit.
The QIC is inherently different. Since the circuit acts only on the target qubit, there has to be an output for any configuration $\tilde{x}$ of the input qubits, even if $\tilde{x}$ is not present in the target distribution.
The unseen data $\tilde{x}$ is expressed as a hole in the probability distribution, which corresponds to the values $p_{\tilde{x},0}=p_{\tilde{x},1}=0$. 
The contribution in the Battacharrya coefficient in \eqref{eq:Hellinger} for the unseen data is null, and the training tends to optimize with respect to the seen data. 

Because of the normalization of the probability distribution output of the QIC ($\sum_xp_x=1$), the distance $d_H$ cannot be 0, even in the case of optimal reproduction of the training dataset. In order to encompass this issue, we calculate $d_H$ only on the support provided by the seen data. This allows the minimum of $d_H$ to be 0, a condition that would reflect a perfect reproduction by the circuit of the distribution of the seen data.
In the following we will show $d_H$ calculated on the probability distribution with partial (seen) data  $p^\text{p}$, and with new (unseen) data $p^\text{n}$, both renormalised such that $\sum_xp^\text{p}_x=\sum_xp^\text{n}_x=1$.
Note that this condition represents another important difference between the QIC and the QCBM: even when the Hellinger distance vanishes, the circuit is able to generalize to unseen data, as we are going to show next.

\subsubsection{Numerical simulations}

\begin{figure}
\centering
 \includegraphics[width=.31\textwidth]{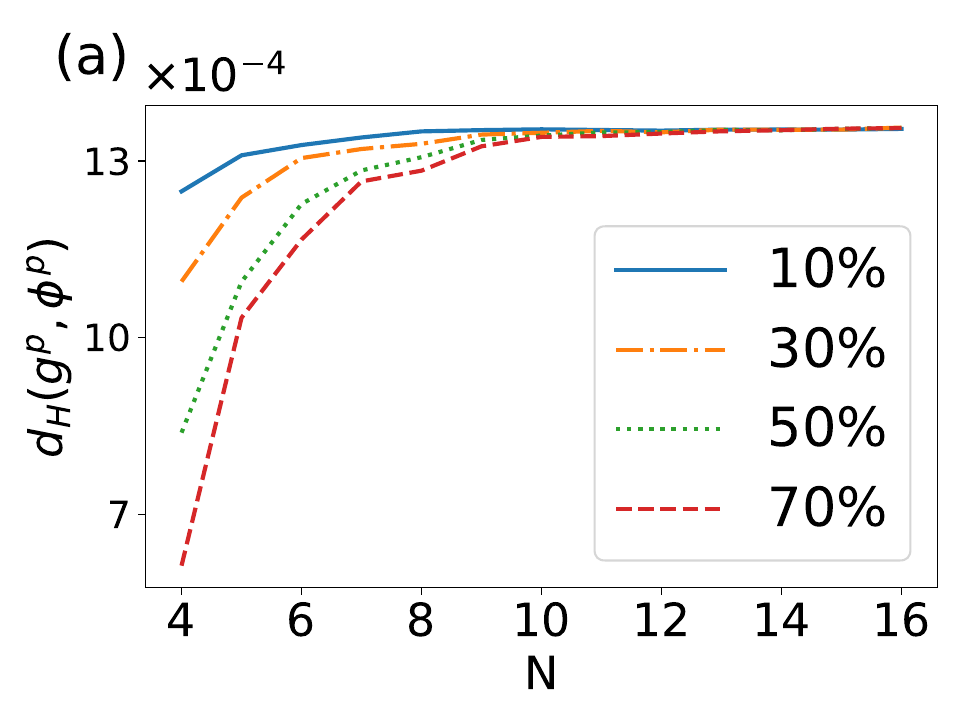}
   \includegraphics[width=.31\textwidth]{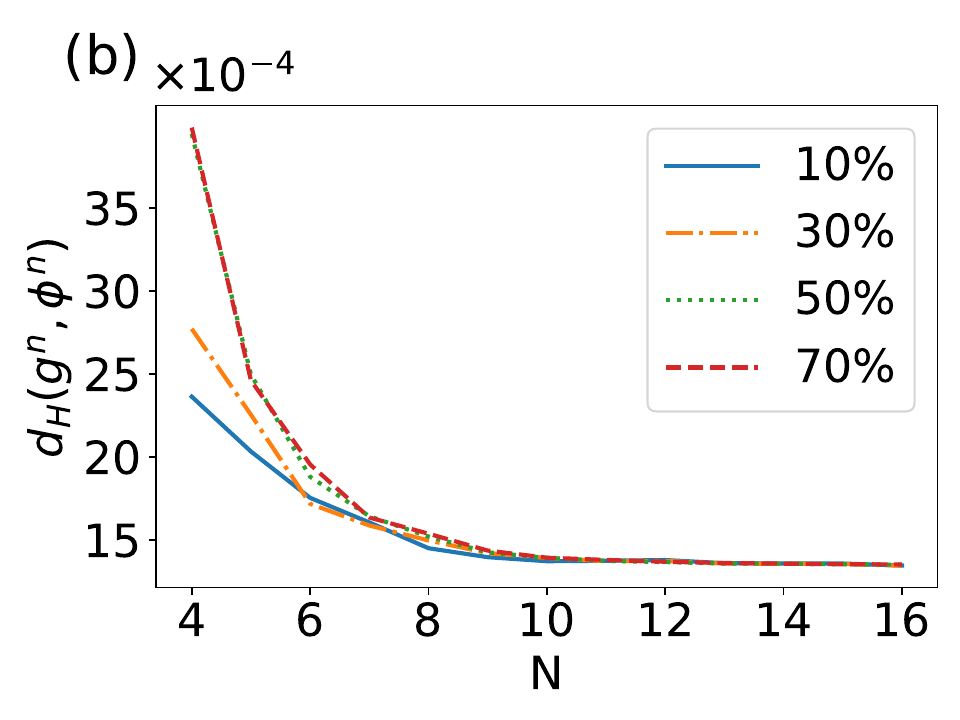}
     \includegraphics[width=.31\textwidth]{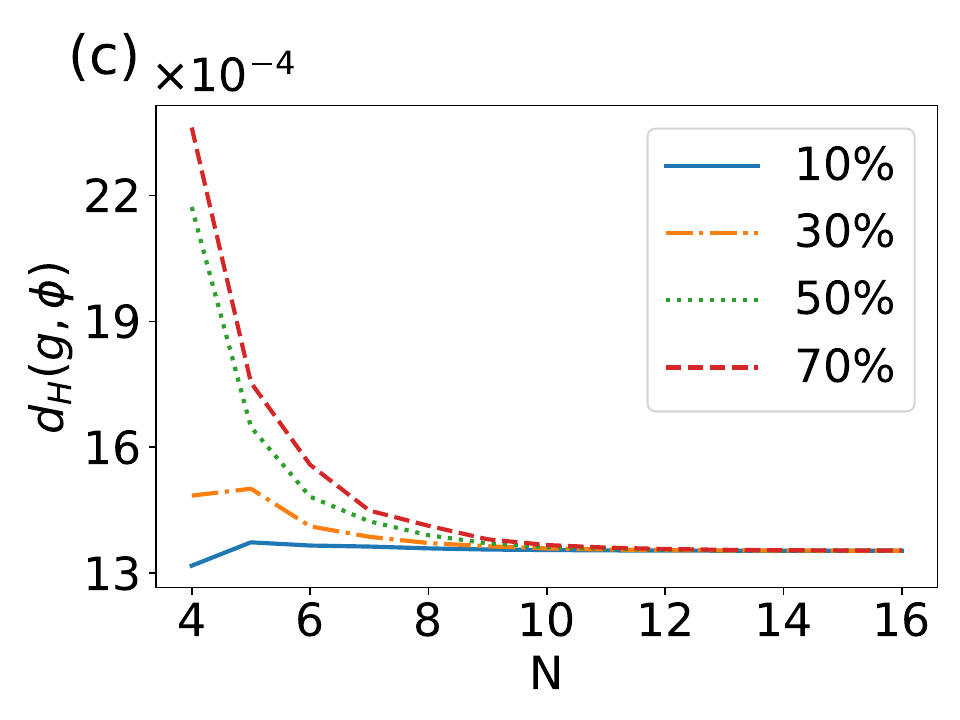}
\caption{The generalization capability of the QIC for the Gaussian distribution. The Hellinger distance between (a) the partial distribution $g^\text{p}$ and the corresponding output distribution $\phi^{\text{p}}$, (b) the distribution of the unseen data $g^\text{n}$ and the corresponding output distribution $\phi^{\text{n}}$, and (c) the complete distribution $g$ and the optimized output distribution $\phi$. Different coloured curves correspond to different precentages of missing data.} 
   \label{fig:figure6}
\end{figure}

In order to understand if the QIC is able to generalize, we subtract from the target distribution a certain percentage of data ($10\%,30\%,50\%$ and $70\%$) and optimize the QIC with respect to the partial distribution. 
Fig.~\ref{fig:figure6} shows the $d_H$ obtained for the Gaussian distribution using the linear QIC. The results we have obtained for the quadratic circuit are similar up to $10^{-2}$. In Fig.~\ref{fig:figure6}(a) we see that the distance between the partial Gaussian distribution $g^\text{p}$ and the partial output distribution $\phi^\text{p}$ tends to a constant value for $N>10$.  This is reflected in the Hellinger distance between the unseen Gaussian distribution $g^\text{n}$ and the new output distribution $\phi^\text{n}$, plotted in Fig.~\ref{fig:figure6}(b), where the distance decreases for larger $N$, till it reaches a plateau. 
Fig.~\ref{fig:figure6}(c) shows the distance of the outcome distribution $\phi$ with the Gaussian distribution $g$.

The different curves converge to the same value for larger $N$ (Fig.~\ref{fig:figure6}) and it is due to the fact that increasing the number of qubits, the dataset becomes large enough that even losing $70\%$ of it the algorithm can still find the pattern in the data to reconstruct the distribution. 

We see therefore that for a large number of $N$ the QIC is able to produce the distribution of dataset and generalize to unseen data.

\begin{figure}
\centering
 \includegraphics[width=.31\textwidth]{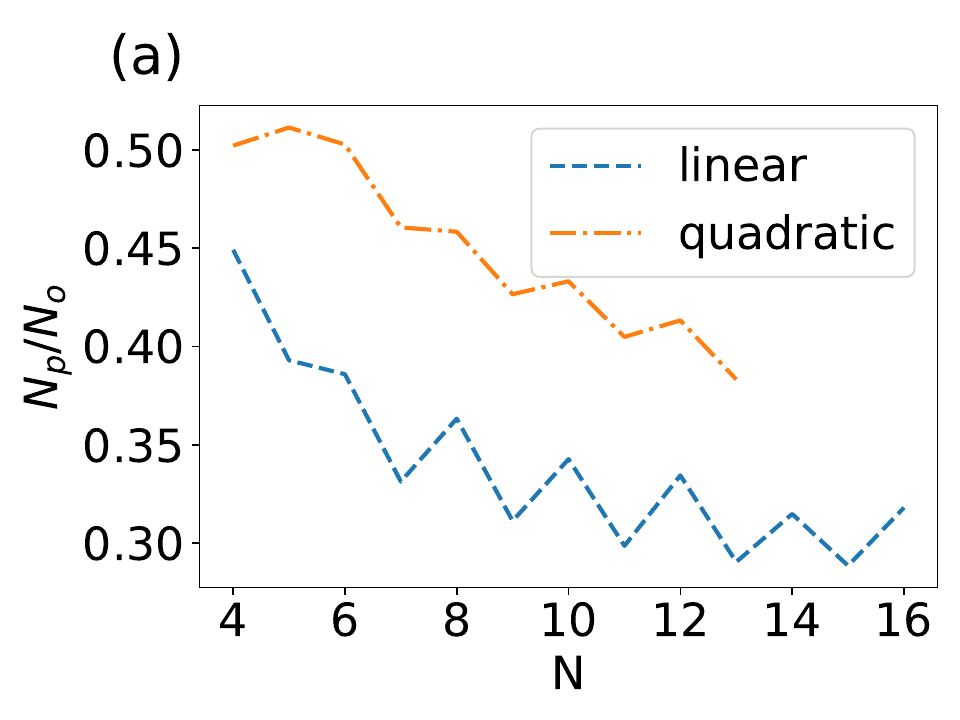} 
 \includegraphics[width=.31\textwidth]{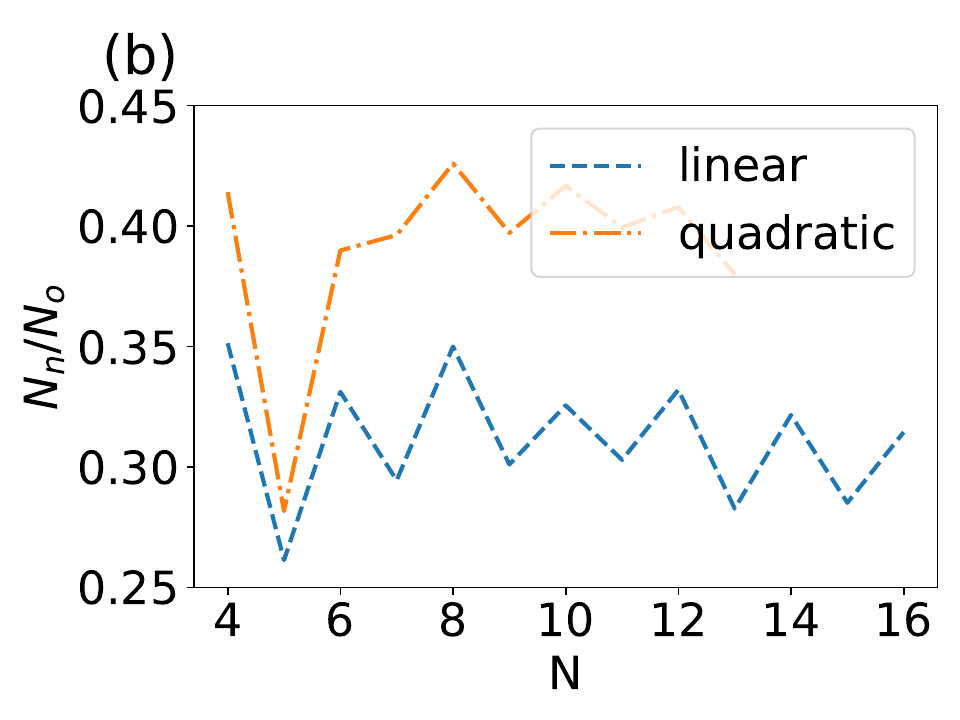} 
 \includegraphics[width=.31\textwidth]{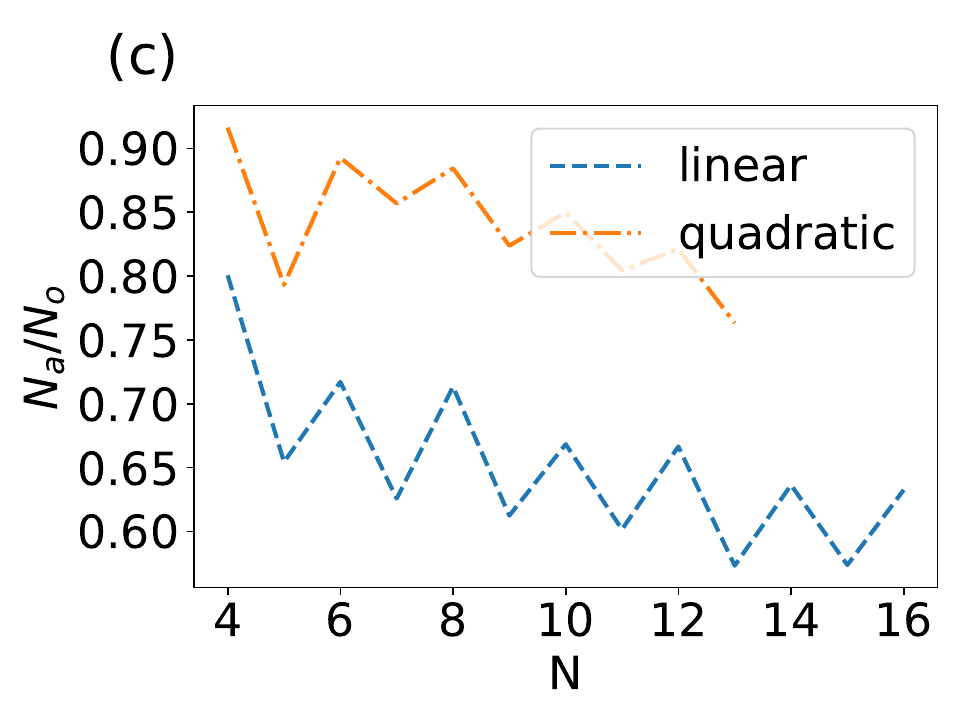} 
\caption{The generalization capability of the QIC for the Majority distribution $m$. In the figure we show the ratio of the correct outcomes from $N_o=1024$ random extraction from the optimized random distribution $\phi$ in the case of a missing portion of $70\%$ . In (a) the ratio $N_p/N_o$ of extractions that belong to partial distribution $m^\text{p}$, in (b) ratio $N_n/N_o$ of extractions that belong to the unseen distribution $m^\text{n}$, and in (c) the ratio $N_a/N_o$ of correct or extractions from the original distribution $m$.} 
   \label{fig:figure7}
\end{figure}

To analyse the behavior of the circuit with the majority distribution $m$ for a partial vision of the data, we choose a different approach.
When for a given input string $n$ there is only one possible output $x$, i.e. $p(n|x)=1$, as it happens for the distribution $m$, we can describe the generalization capability of the QICs in terms of the number of correct strings that we obtain as output of the circuit. We define, out of $N_\text{o}$ random outcomes, the number of strings that belong to $m^\text{p}$ as $N_\text{p}$, and the one that belong to $m^\text{n}$ as $N_{\text{n}}$.  In Fig.~\ref{fig:figure7} we plot the results obtained when the portion of missing data is $70\%$. For this distribution the results obtained by the linear and quadratic circuits are very different, since the latter is more able to reproduce the original distribution $m$. We plot the ratios $N_\text{p}/N_\text{o}$,  $N_\text{n}/N_\text{o}$ and the ratio  of acceptable outcomes $N_\text{a}/N_\text{o}$ with $N_\text{a}=N_\text{p}+N_\text{n}$, in Fig.~\ref{fig:figure7} (a), (b) and (c) respectively. 
The results show that even with a $70\%$ of unseen data, about $90\%$ of the outcomes followed the original rule in the case of $2^4$ input data, and about $70\%$ of them in the case of $2^{16}$ input data. 

\section{Conclusions}\label{sec:V}

In this work, we have explored the possibility of using a quantum computer to impute a missing attribute of a data point, given a statistical distribution of the attribute within a dataset. We have given a brief introduction to the most commonly used classical techniques, whether they are based on statistical inference or machine learning algorithms. Consequently, inspired by the Quantum Born Machine, we have proposed a quantum circuit, the QIC, for imputing classical data by employing parametric gates.

In the first part of the paper we introduced the QIC algorithm. We calculated the theoretical upper bound of the loss function used for optimization depending on the number of parameters of the algorithm. This limit can be used as a red flag. If the final Hellinger distance exceeds this upper bound, it signals issues within the optimization process. 

Since the QIC has a relatively simple structure, we were able to find the analytical solution of the optimal angles. 
However, our circuit was still able to develop entanglement. We have tested this feature by calculating the entanglement entropy between the input qubits and the target qubit, and by verifying the presence of BPs, that are characteristic of circuits with entanglement. Furthermore our method avoids the problem of sampling the output of the quantum circuit.
We acknowledge that this is not a standard procedure. However, by leveraging the particular circuit ansatz we were able to avoid the optimization through successive measurements.

In the second part of the manuscript, we have tested our circuit to reproduce several types of probability distributions, finding the the optimization goes well below the upper bound limit, reaching values of the Hellinger distance in the order of $10^{-2}$ even for randomly generated distributions. 

Finally, we have addressed the ability of the QIC to generalize to unseen data in the dataset, i.e. we questioned if, given an implicit rule, the output of the optimized QIC would follow that rule even for data points that did not belong to the training set. We have tested the QIC on the Gaussian and the majority distributions with an increasing portion of missing data. We have found that that algorithm was able to recover the true value of the missing data even when the available data were a small fraction of the dataset. 

We believe these results can lead the way to the use of quantum circuits for the imputation of classical data. 

\section{Acknowledgements}
This research is funded by the International Foundation Big Data and Artificial Intelligence for Human Development (IFAB) through the project “Quantum Computing for Applications”. E. E. is partially supported by INFN through the project “QUANTUM” and the PRIN2022H77XB7 project. C. S. and E. E. acknowledge financial support from the National Centre for HPC, Big Data and Quantum Computing (Spoke 10, CN00000013).

\bibliography{mybib}

\appendix

\section{Barren Plateaus in the QIC}\label{sec:appI}

In this appendix we analyze the emergence of Barren plateaus (BPs) in the circuit. BPs are large portions of the parameter space where the gradient of the cost function $\partial_{\theta_i}C$ for all the different parameters $\theta_i$ is zero. The BP are the consequence of the concentration of measure in the exponentially increasing volume of the Hilbert space of $N$ qubits~\cite{Mcclean2018}. In such spaces, the variance of the gradient decays exponentially with the number of qubits, as $\langle(\partial_{\theta_i}C)^2\rangle\sim2^{-N}$. 
This kind of behavior is an obvious problem for gradient based optimization methods, but it can raise issues also for other global optimization methods, as the Bayesian optimization strategy we adopt in this analysis. In order to avoid the BPs, different strategies have been explored in several papers~\cite{Patti2021,Wiersema2021}, but in summary they all convey that the BPs emerge when the system is subjected to large entanglement, either in the circuit~\cite{Ortiz2021,Patti2021} or in the definition of the cost function~\cite{Cerezo2021,Uvarov2021,Garcia2023}. 
We found this behavior also in our circuit.

In fact, the presence of BPs in the parameter space highly affects the ability of the optimization algorithm of finding the global minimum. 

\begin{figure}
\centering
 \includegraphics[width=.4\textwidth]{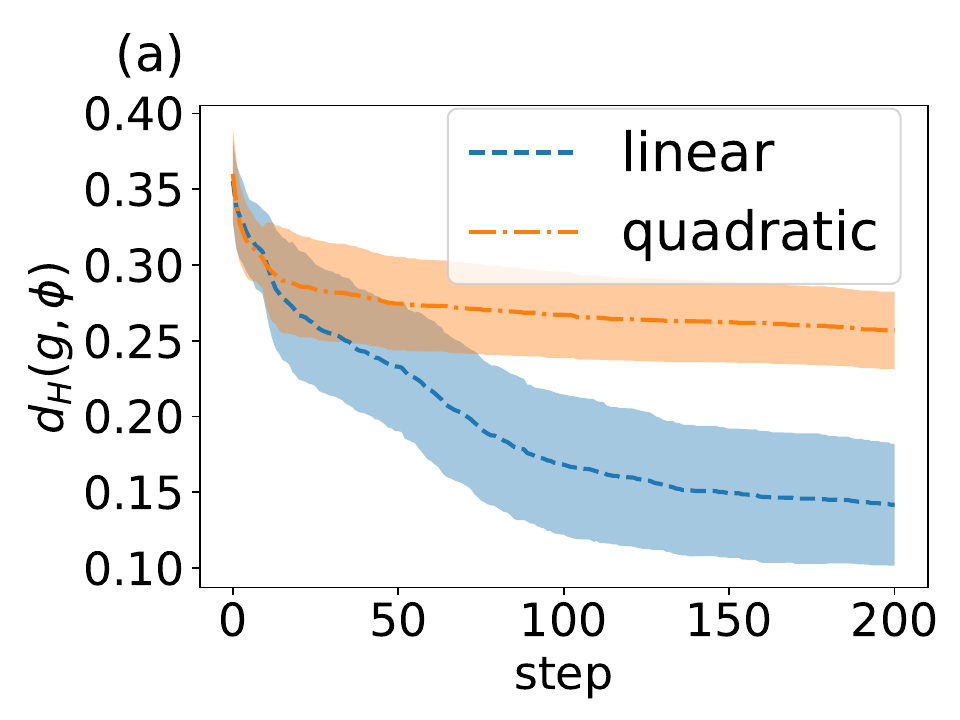}
 \includegraphics[width=.4\textwidth]{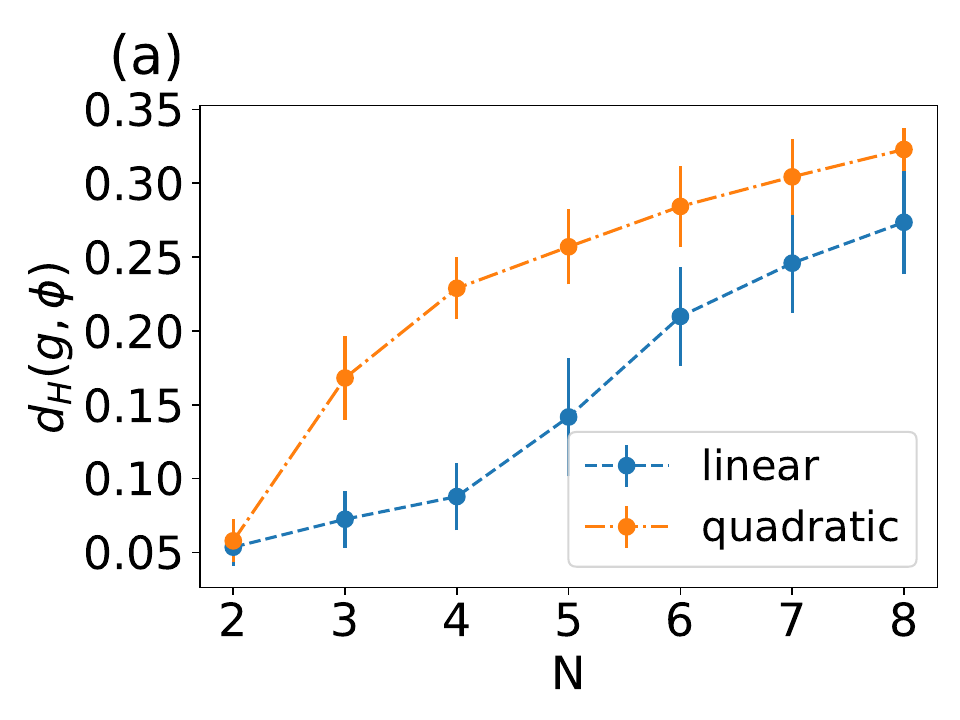}
\caption{The training of the Gaussian distribution using measurements for the linear and the quadratic circuit. In (a) the distance $d_H$ at different steps of the Bayesian optimization algorithm. In (b) the optimal value of $d_H$ varying the number of input qubits $N$.} 
   \label{fig:fig_app1}
\end{figure}

\noindent Fig.~\ref{fig:fig_app1}(a) shows that the Hellinger distance reached by the Bayesian optimization algorithm at a certain step. We stop the optimization algorithm after a number of steps increasing with the number of qubits, with $t_{\max}=100+50(N-3)\times\max(3,N)$. Fig.~\ref{fig:fig_app1}(b) shows the value of the Hellinger distance found after the optimization, varying the number of input qubits $N$.
Note that in Figs.~\ref{fig:fig_app1}(a) and (b) the target distribution is the Gaussian distribution, but analogue results have been obtained for the other tested distributions. 

\begin{figure}[ht!]
\centering
\includegraphics[width=.3\textwidth]{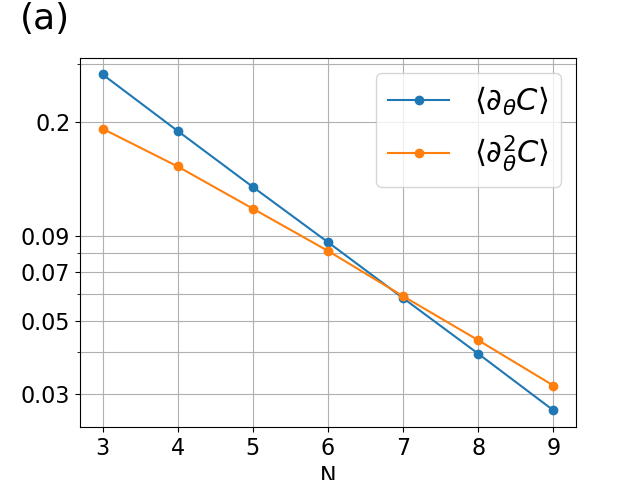}
\includegraphics[width=.3\textwidth]{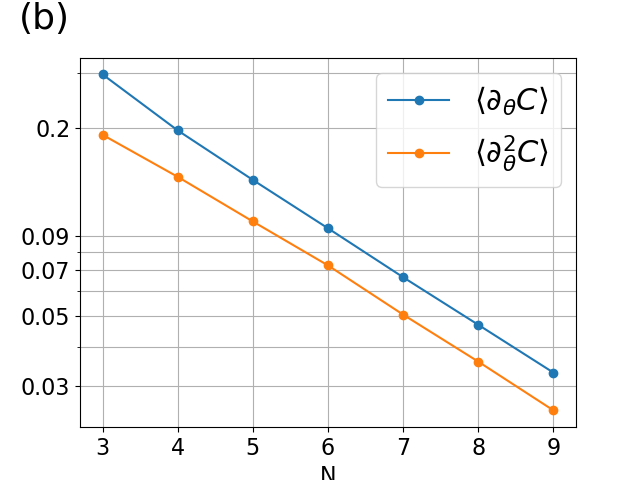}
\includegraphics[width=.3\textwidth]{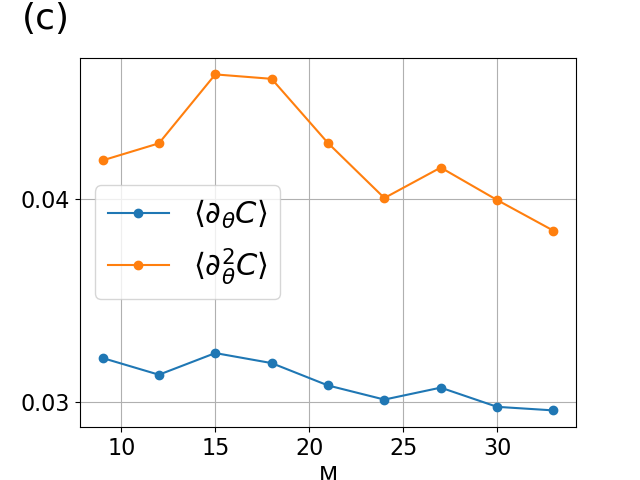}
\caption{The mean value and the variance of the gradient of the cost function in the parameter space varying the number of qubits $N$ for (a) the linear circuit, and for (b) the quadratic circuit. In (c) the mean value and the variance of the gradient of the cost function varying the number of parameters $M$, with fixed $N=9$. \label{fig:fig_app2}}
\end{figure}

In Fig.~\ref{fig:fig_app2} we show the mean value and the variance of the gradient of the cost function in the parameter space for (a) the linear circuit, where the number of parameters scales as $N$ and (b) the quadratic circuit, where the number of parameters scales as $N^2$. In both cases we see the exponential decrease of the variance and of the mean value, in agreement with the expectation.   

In Fig.~\ref{fig:fig_app2}(c) we show how the mean value and the variance of the gradient vary when the number of parameters is increased. 
In order to do so, we fix the number of qubits to $N=9$ and we add $C_2NOT$ gates to the linear circuit until it becomes the full quadratic circuit. 
Contrary to the behaviour obtained for Figs.~\ref{fig:fig_app2}(a),(b) those curve do not follow an exponential trend. We can explain this behaviour by analysing the entanglement in the circuit.

In fact, the emergence of BPs in the parameter space is related to the presence of entanglement in the circuit. 
We calculate the entanglement entropy  $S=-\text{Tr}[\rho_t\log\rho_t]$, on the state of the target qubit $\rho_t$, obtained tracing out the $N$ input qubits. $S$ quantifies the entanglement  between the target qubit and the rest of the circuit. 
In order to relate the entanglement of the circuit to the landscape on the parameter space, we average $S$ over the volume $V_M$ of the $M$ parameters $\alpha_1,\dots,\alpha_M$, yielding 

\begin{equation}
\bar{S} = \frac{1}{V_M}\int_{V_M}d^M\alpha S(\alpha_1,\dots,\alpha_M),
\end{equation}

\noindent that is the expectation value of $S$ when we run the circuit with a random choice of the parameters. 

\begin{figure}[ht!]
\centering
\includegraphics[width=.4\textwidth]{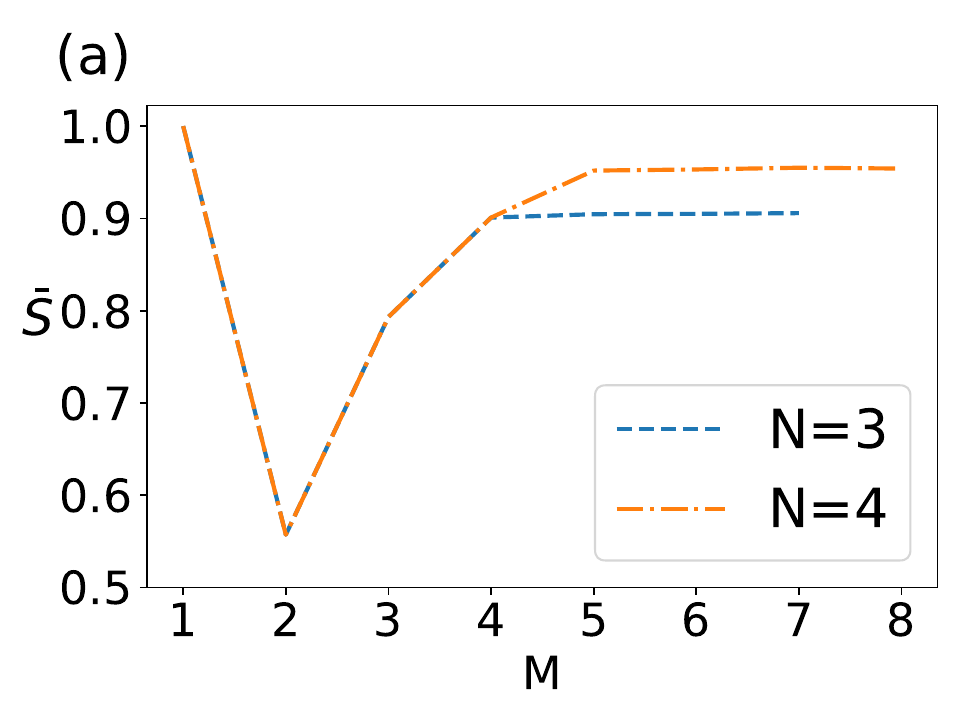}
\includegraphics[width=.4\textwidth]{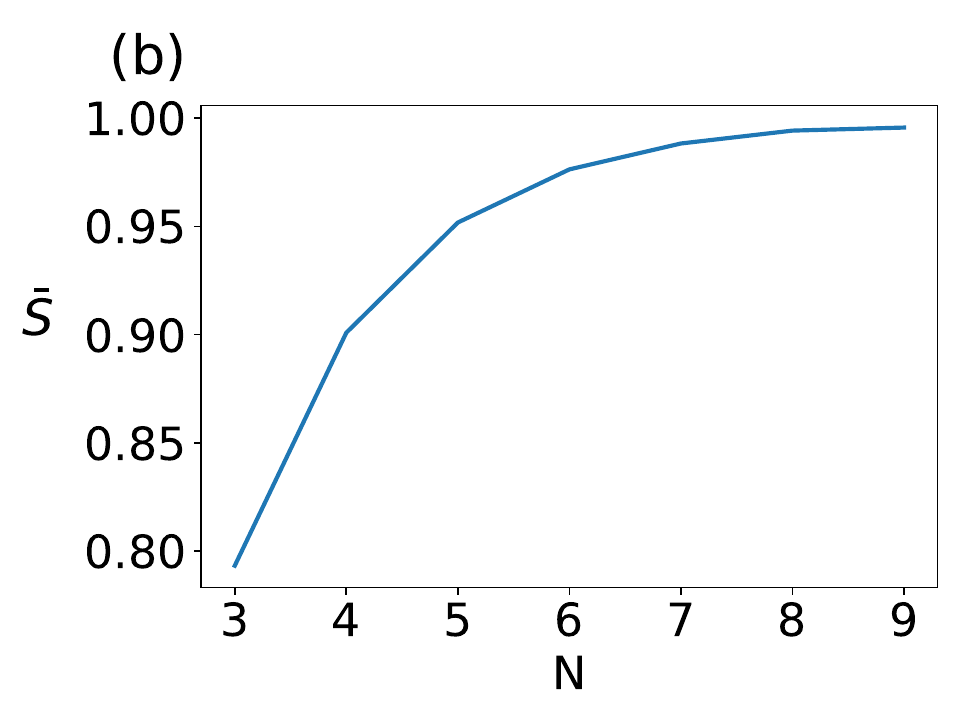}
\caption{(a) The mean value of the entanglement entropy $\bar{S}$ as a function of the number of parameters of the QIC, at fixed number of input qubits. (b) the mean value $\bar{S}$ obtained by the linear QIC varying the number of qubits.     \label{fig:fig_app3}}
\end{figure}

In Fig.~\ref{fig:fig_app3}(a) we plot $\bar{S}$ for $N=3$ and $N=4$  as a function of the number of parameters $M$. The entanglement increases until it reaches a plateau, when the number of $\text{CNOT}$s (and therefore the number of parameters $M$) is the same as the number of qubits. When the number of parameters exceeds the number of qubits, we introduce $\text{C}_2\text{NOT}$s in the circuit. We see that the addition of the $\text{C}_2\text{NOT}$s doesn't change the level of entanglement in the circuit, and this is in accord with the curves in Fig.~\ref{fig:fig_app2}(c). In Fig.~\ref{fig:fig_app3}(b) we consider the linear circuit, with $M=N+1$ and we plot the mean value of the entanglement entropy as a function of the number of qubits. In this case, we have found $\bar{S}\sim 1 - a^{-b(N - c)}$, with $a\approx 1.2, b\approx 3.2, c\approx 0.8$, thus following an exponential function of $N$. 


\section{Analytical Solution}\label{sec:appII}
In this appendix we show an example of matrix form of the unitary of the linear circuit and then derive the formula for the quadratic and exponential cases.

\subsection{Linear Circuit}
In the main text we derived Eq.~\eqref{eq:blocks} for the linear circuit case. Let us as an example plot the matrix form for a circuit with $N=3$ imputation qubits.
\paragraph{Example} The matrix of the circuit with 2 input qubits and one imputation qubit is:
\begin{equation}
    U = 
    \begin{pmatrix}
    R_y\big[\theta_0 + \theta_1 + \theta_2\big] & \boldsymbol{0} & \boldsymbol{0} & \boldsymbol{0} \\
    \boldsymbol{0} & XR_y\big[\theta_0 - \theta_1 -\theta_2\big] & \boldsymbol{0} & \boldsymbol{0} \\
    \boldsymbol{0} & \boldsymbol{0}  &  XR_y\big[\theta_0 + \theta_1 - \theta_2\big]& \boldsymbol{0} \\
    \boldsymbol{0} & \boldsymbol{0} & \boldsymbol{0} & R_y\big[\theta_0 - \theta_1 + \theta_2\big]
    
    \end{pmatrix},\nonumber
\end{equation}
where
\begin{equation}
    R_y(\theta) = 
    \begin{pmatrix}
        \cos \theta & -\sin \theta \\
        \sin \theta & \cos \theta
    \end{pmatrix}\nonumber
\end{equation}
and
\begin{equation}
    XR_y(\theta) = 
    \begin{pmatrix}
        \sin \theta & \cos \theta \\
        \cos \theta & -\sin \theta
    \end{pmatrix}.\nonumber
\end{equation}

\subsection{Quadratic circuit}
\begin{figure}[h!]
  \centering
\begin{quantikz}
\lstick{$q_3$ $\ket{0}$} &\gate{H}\gategroup[3,steps=1,style={dashed,rounded corners,inner xsep=2pt},background]{}              
&\gategroup[4,steps=2,style={inner sep=1pt},background]{{\sc linear}} \dots &\qw  &\qw &\qw &\ctrl{2} &\qw   &\ctrl{1}  &\qw &\gategroup[4,steps=-6,style={inner sep=1pt},background]{{\sc quadratic}}\qw\\
\lstick{$q_2$ $\ket{0}$}&\gate{H}
   &\dots   &\qw  &\ctrl{1} &\qw	&\qw & \qw &\ctrl{2} &\qw &\qw\\
\lstick{$q_1$ $\ket{0}$}&\gate{H}
    &\dots   &\qw  &\ctrl{1} &\qw &\ctrl{1} &\qw &\qw &\qw &\qw\\
\lstick{$q_0$ $\ket{0}$}&\qw
&\dots &\gate{R_y(\theta_3)}&\targ{} &\gate{R_y(\alpha_{1,2})} &\qw &\gate{R_y(\alpha_{1,3})}&\targ{} &\gate{R_y(\alpha_{2,3})}&\meter{}
\end{quantikz} 
\caption{Example of a quadratic circuit with $N=3$ input qubits.}
\label{fig:quadratic_circuit}
\end{figure}
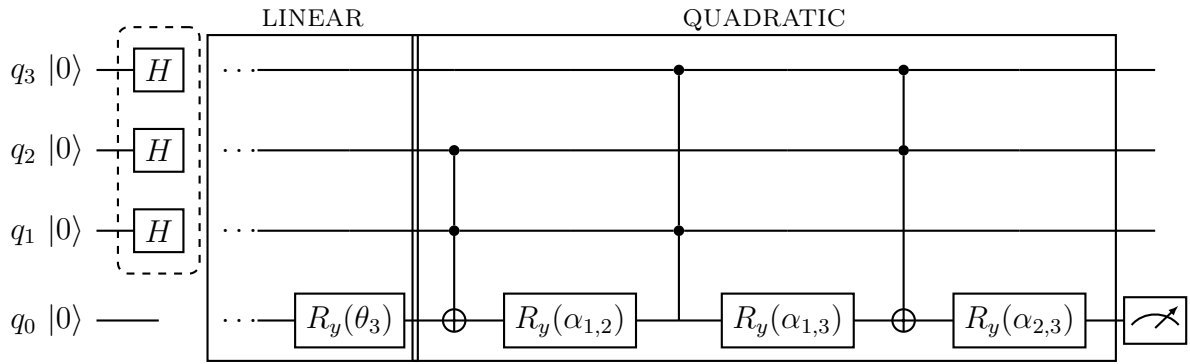

The quadratic circuit introduces a series of Toffoli gates that raises the total parameters (and rotations) to a number proportional to the square of the input qubits. A quadratic circuit is composed of an initial linear part, equivalent to the last section, plus the set of parametric Toffoli gates as shown in Fig.~\ref{fig:quadratic_circuit}.

Notice that the second set of angles is named $\alpha_{i,j}$ where $i,j$ represent the set of control qubits in the preceding Toffoli gate.

As done before, to understand the shape of the total unitary we shift every rotation to the beginning of the circuit. Starting with $R_y(\alpha_{1,2})$ we notice that the angle acquires a phase $-1$ only if $b_1 b_2 = 1 \mod 2$, that is $(-1)^{b_1 b_2}$. The rotation $R_y(\alpha_{1,3})$ consequently acquires a phase $(-1)^{b_1 b_2 + b_1 b_3}$ that depends on both Toffoli gates preceding it. The same goes for the third rotation.
We can simplify the exponent by defining:
\begin{equation}
    Q_{n,m} (\boldsymbol{b}) = \sum_{i=1}^n \sum_{j=i+1}^m b_i b_j \mod 2
\end{equation}
which is the exponent of the phase acquired by the angles when commuting with the Toffoli gates. In this way we obtain the rule
\begin{equation}
    R_y(\boldsymbol{\alpha}) = R_y \bigg[\sum_{n = 1}^{N-1} \sum_{m = n+1}^N (-1)^{Q_{n,m}} \alpha_{n,m} \bigg]
\end{equation}
where $\boldsymbol{\alpha}$ represents the sum of all the angles.
The same rule applies to the $X$ rotation enforced by the Toffoli gates, so the total action of only the quadratic part gives us block matrices like such:
\begin{equation}
    U_{\boldsymbol{q}} = X^{Q_{N-1, N}} R_y \bigg[\sum_{n = 1}^{N-1} \sum_{m = n+1}^N (-1)^{Q_{n,m}} \alpha_{n,m} \bigg]
\end{equation}
Now, since these angles also need to commute with the CNOTs, in addition to $Q_{n,m}$ we need to add the contribution from $S_n$ to the phase.
Putting it together with the results of the linear part (eq.~\eqref{eq:blocks}) we obtain
\begin{eqnarray}
    U_{\boldsymbol{q}} = X^{S_N + Q_{N-1, N}} \times R_y \Bigg[ \sum_{n=1}^N \bigg[(-1)^{S_n} \theta_n + \sum_{m = n+1}^N (-1)^{Q_{n,m} + S_n} \alpha_{n,m} \bigg]\Bigg]
\end{eqnarray}

\subsection{Exponential circuit}
We define the exponential circuit as the circuit composed of the linear circuit plus all possible combination of n-Toffoli gates with $n = 2, \dots N$. Notice that, with this nomenclature, the quadratic is the sub-circuit of the exponential circuit with $n = 2$.

To unify and simplify notation we define the exponential phase:
\begin{eqnarray}
    E_{J} (n_1, \dots, n_J; \boldsymbol{q})
    \equiv \sum_{a_1 = 1}^{n_1} \sum_{a_2 = a_1 +1}^{n_2} \dots
    \sum_{a_J = a_{J-1} + 1}^{n_J} b_{a_1}b_{a_2} \dots b_{a_J} \mod 2
\end{eqnarray}
so that we can regain the two previous phases:
\begin{eqnarray}
    E_1(n_1; \boldsymbol{b}) &= \sum_{a_1 = 1}^{n_1} b_{a_1} \equiv S_{n_1} \\
    E_2 (n_1, n_2; \boldsymbol{b}) &= \sum_{a_1 = 1}^{n_1} \sum_{a_2 = a_1 + 1}^{n_2} b_{a_1} b_{a_2} \equiv Q_{n_1, n_2}
\end{eqnarray}
In this way, we can define the $(2\times 2)$ matrix blocks with the recursive formula:
\begin{eqnarray}
U_{\boldsymbol{q}} &=& X^{E_1(N) + E_2(N-1, N) + \dots E_N(1, \dots, N)} \nonumber\\
&&\times R_y\Bigg[ \sum_{n_1=1}^N \bigg[(-1)^{E_1} \theta_{n_1} + \sum_{n_2 = n_1 + 1}^N (-1)^{E_2} \alpha_{n_1,n_2} + \big[ \dots \nonumber\\
&&+\sum_{n_N = n_{N-1} + 1}^N (-1)^{E_N} \alpha_{n_1,\dots,n_N}\big] \dots \bigg]\Bigg]
\end{eqnarray}

\end{document}